\documentclass[conference]{IEEEtran} 

\usepackage[utf8x]{inputenc} 
\usepackage{graphicx}
\usepackage{mathtools, amsmath,amssymb}
\usepackage[ruled,vlined]{algorithm2e}
\usepackage{caption, subcaption}
\usepackage{todonotes}
\usepackage[para]{footmisc}
\usepackage{booktabs}
\usepackage{multirow}
\usepackage{xcolor,colortbl}
\usepackage{hyperref}
\usepackage{cleveref}
\usepackage[noadjust]{cite}

\def\BibTeX{{\rm B\kern-.05em{\sc i\kern-.025em b}\kern-.08em
    T\kern-.1667em\lower.7ex\hbox{E}\kern-.125emX}}
    
\begin{document} % 

\title{OncoNetExplainer: Explainable Predictions of Cancer Types Based on Gene Expression Data}

\author{
\IEEEauthorblockN{
	Md. Rezaul Karim\IEEEauthorrefmark{1}\IEEEauthorrefmark{2},
	Michael Cochez\IEEEauthorrefmark{3}, 
	Oya Beyan\IEEEauthorrefmark{2}\IEEEauthorrefmark{1},
	Stefan Decker\IEEEauthorrefmark{2}\IEEEauthorrefmark{1},
	and Christoph Lange\IEEEauthorrefmark{1}\IEEEauthorrefmark{2}
}
\IEEEauthorblockA{
	\IEEEauthorrefmark{1} Fraunhofer Institute for Applied Information Technology FIT, Germany  
}
\IEEEauthorblockA{
	\IEEEauthorrefmark{2} Information Systems and Databases, RWTH Aachen University, Germany
}
\IEEEauthorblockA{
	\IEEEauthorrefmark{3} Department of Computer Science, Vrije Universiteit Amsterdam, the Netherlands
}
}

%\author{
%\IEEEauthorblockN{Md. Rezaul Karim}
%\IEEEauthorblockA{Fraunhofer FIT, Germany} 
%\IEEEauthorblockA{RWTH Aachen University, Germany}
%\and
%\IEEEauthorblockN{Michael Cochez}
%\IEEEauthorblockA{} \\
%\textit{RWTH Aachen University, Germany}
%%\and
%%\IEEEauthorblockN{Giuseppe Bonaccorso}
%%\IEEEauthorblockA{\textit{Fraunhofer FIT, Germany} \\
%%\textit{RWTH Aachen University, Germany}}
%%\and
%%\IEEEauthorblockN{Ivan G. Costa}
%%\IEEEauthorblockA{\textit{Fraunhofer FIT, Germany} \\
%%\textit{RWTH Aachen University, Germany}}
%\and
%\IEEEauthorblockN{Oya Beyan}
%\IEEEauthorblockA{\textit{RWTH Aachen University, Germany} \\
%\textit{Fraunhofer FIT, Germany}}
%\and
%\IEEEauthorblockN{Christoph Lange}
%\IEEEauthorblockA{\textit{Fraunhofer FIT, Germany} \\
%\textit{RWTH Aachen University, Germany}}
%\and
%\IEEEauthorblockN{Stefan Decker}
%\IEEEauthorblockA{\textit{Fraunhofer FIT, Germany} \\
%\textit{RWTH Aachen University, Germany}}
%\\[-20.0ex]}

\maketitle

\begin{abstract}
The discovery of important biomarkers is a significant step towards understanding the molecular mechanisms of carcinogenesis; enabling accurate diagnosis for, and prognosis of, a certain cancer type.
Before recommending any diagnosis, genomics data such as gene expressions~(GE) and clinical outcomes need to be analyzed.
However, complex nature, high dimensionality, and heterogeneity in genomics data make the overall analysis challenging.
Convolutional neural networks~(CNN) have shown tremendous success in solving such problems.
However, neural network models are perceived mostly as `black box' methods because of their not well-understood internal functioning.
However, interpretability is important to provide insights on why a given cancer case has a certain type.
Besides, finding the most important biomarkers can help in recommending more accurate treatments and drug repositioning.
Moreover, the `right to explanation' of the EU GDPR gives patients the right to know why and how an algorithm made a diagnosis decision.
Hence, in this paper, we propose a new approach called \emph{OncoNetExplainer} to make explainable predictions of cancer types based on GE data.
We used genomics data about 9,074 cancer patients covering 33 different cancer types from the Pan-Cancer Atlas on which we trained CNN and VGG16 networks using guided-gradient class activation maps++~(GradCAM++).
Further, we generate class-specific heat maps to identify significant biomarkers and computed feature importance in terms of mean absolute impact to rank top genes across all the cancer types.
Quantitative and qualitative analyses show that both models exhibit high confidence at predicting the cancer types correctly giving an average precision of 96.25\%.
To provide comparisons with the baselines, we identified top genes, and cancer-specific driver genes using gradient boosted trees and SHapley Additive exPlanations~(SHAP). 
Finally, our findings were validated with the annotations provided by the TumorPortal.
\end{abstract}

\section{Introduction}
\label{section1}
Cancer is caused by gene alterations and abnormal behaviors of genes that control cell division and cell growth. The change in the structure of occurring genetic aberrations, such as somatic mutations, copy number variations~(CNV), profiles, and different epigenetic alterations are unique for each type of cancer~\cite{82Tomczak,19Cruz}. As a result, gene expressions~(GE) can be disrupted by cell division or environmental effects, or genetically inherited from parents. Changes in GE sometimes change the production of different proteins, affecting normal cell behavior. These damaged cells start reproducing more rapidly than usual and gradually increase in the affected area by forming a tumor. Intermittently, such tumors turn into a type of cancer~\cite{zuo2019identification,24Podolsky}. This is one of the utmost reasons cancer incidences are gradually increasing every year and have become the second leading cause of death worldwide. 
Consequently, more than 200 types of cancer have been identified, each of which can be characterized by different molecular profiles requiring unique therapeutic strategies~\cite{82Tomczak}. 
%According to a statistics from the National Cancer Institute
%\footnote{\url{https://www.cancer.gov/about-cancer/understanding/statistics}} 
%observed 14.1 million deathly cancer cases in 2012, out of which 8.8 million people died of five leading cancers of lung, liver, colon/rectum, stomach, and breast~\cite{zuo2019identification}. 
The most common cancers diagnosed in men are prostate, lung, and colorectal cancers, while for women, breast, lung, and colorectal cancers are most common~\cite{li2017comprehensive}.

%In 2018, an estimated 17.35 million cancer cases were diagnosed in the United States in which 609,640 people died. The number of new cancer cases per year is expected to rise to 23.6 million by 2030, which is anticipated to increase further to 70\% by 2035~\cite{71Torre}. 
As the importance of genetic knowledge in cancer treatment is increasingly addressed, several projects have emerged %recently
, of which The Cancer Genome Atlas~(TCGA) most well-known for omics data. 
%Omics data is a collection of biomolecules inside living things such as genomics, metabolomics, and proteomics. 
TCGA curates various omics data, e.g., gene mutation, gene expressions (GE), DNA methylation, CNV, and miRNA expressions. By acquiring deep insights of these data, treatment can be focused on preventive measures. Besides, clinical outcomes, i.e., clinical and pathology information are provided. TCGA further analyzed over 11,000 cancer cases from 33 prevalent forms of cancer, which fostered the accomplishment of the Pan-Cancer Atlas~(PCA), which results from the normalized GE data about 20K protein-coding genes~\cite{lyu2018deep}. 

These data, however, are highly variable, high-dimensional, and sourced from heterogeneous platforms, which imposes significant challenges to existing bioinformatics tools stimulating the development of deep learning~(DL)-based diagnosis and prognosis systems. Since DL algorithms work better with such high dimensional data, recent studies focused on using deep architectures such as autoencoders, CNN, and Recurrent Neural Networks~(RNN). Although these models have shown tremendous success in exhibiting high confidence, they are mostly perceived as `black box' methods because of a lack of understanding of their internal functioning. 
This is a serious drawback since interpretability is essential to generate insights on why a given cancer case is of a certain type, and since knowing the most important biomarkers can help in recommending more accurate treatments and drug repositioning. Further, the `right to explanation' of the EU GDPR~\cite{kaminski2019right} gives patients the right to know why and how an algorithm made a diagnosis decision. However, existing approaches can neither ensure the diagnosis transparently nor are they trustworthy.

Since GE data are very high dimensional and a significant number of genes have a trivial effect on the tumor making them very weak features, we hypothesize that our approach called \emph{OncoNetExplainer} based on neural networks~(NN) and ML baselines with the explanation capability can be more effective at learning hierarchical features. We trained and evaluated CNN and VGG16 networks with a guided-gradient class activation map~(GradCAM++)~\cite{chattopadhay2018grad}. Using GradCAM++, we generated heat maps~(HM) for all the classes showing prominent pixels across GE values and computed the feature importance in terms of mean absolute impact~(MAI) to identify important biomarkers and provide interpretations of the predictions to make the cancer diagnosis transparent. Further, we validated our findings through functional analysis to make sure the selected genes are biologically trustworthy for the corresponding tumor types. Further, SHAP
%, an explainable framework, 
is used along with GBT to validate our findings based on the annotations provided by the TumorPortal to ensure the consistency and accuracy. 

\iffalse
The overall contributions of this paper can be summarized as follows:  
%\vspace{2mm} 
\begin{itemize}
\item Preparation of rich labelled gene expression data across 33 different cancer types. 
\item An effective way of training CNN, identifying the most significant gene biomarkers using GradCAM and providing class specific explanations giving the top-10 genes for each cancer type and common genes across types 
\item Comprehensive evaluations with detailed analyses of outcomes and comparisons with the state of the art. 
\end{itemize}
%\vspace{2mm} 
\fi 

The rest of the paper is structured as follows: \cref{related_work} discusses related works and highlights their potential limitations. \Cref{dc} chronicles data collection and preparation before constructing and training the network. \Cref{results} demonstrates some experimental results and discusses the key findings of the study. \Cref{conclusion} provides explanations and points out the relevance of the study, highlights its limitations and discusses future works before concluding the paper.

\section{Related work}
\label{related_work}
Numerous approaches using genomic data, bioimaging, and clinical outcomes have been proposed for analyzing genomic profiles of patients for treatment decision making~\cite{li2017comprehensive}. 
However, early detection of tumors is particularly important for better treatment of patients, a notable issue being the discrimination of tumor samples from normal ones~\cite{liu2017tumor}. % by analyzing genomics data.% and decision making for cancer treatment. 
%For example, RNA-Seq data used more widely to identify rare and common transcripts, isoforms, and non-coding RNAs in cancer. %Single nucleotide polymorphism data used to identify segmental variations across multiple cancer genomes, and array-based DNA methylation data is used to provide epigenetic changes in the genome that are useful for early genetic changes and early-stage metabolomic detection of ovarian cancer~\cite{82Tomczak,95Gaul}. 
Unlike conventional cancer typing methods that work based on morphological appearances, GE levels of the tumor are used to differentiate tumors that have similar histo-pathological appearances, giving more accurate tumor typing results for colorectal cancer diagnosis~\cite{paroder2006na+}. Different types of mutation data, e.g., point mutation, single nucleotide variation, INDEL, and CNV are also used. Yuan et al.~\cite{yuan2018cancer} observed that these genomics phenomena are associated with complex diseases and contribute to the growth of different types of cancers. 

Different ML algorithms were trained using mixed data types, e.g., genomic data, bioimaging data, and clinical outcomes. These approaches not only proved useful at improving cancer prognosis, diagnosis, and treatments but also revealed subtype information of several cancer types~\cite{66Huang}. 
Li et al.~\cite{li2017comprehensive} employs a genetic algorithm for feature selection and a k-nearest neighbors for the classification based on GE data from the PCA project. Their approach can classify 90\% of the tumor cases correctly using different 20-gene sets. However, since the data contain GE values of more 20K protein-coding genes, these generic ML methods were found to be inefficient, with some genes appearing repeatedly in the sets because of the curse of dimensionality. 
\iffalse
\begin{table*}[!ht]
\vspace{1mm} 
\caption{Different cancer detection methods, data types, and performance }
\label{table:stateofart}
\begin{center}
\small
\begin{tabular}{llllll}
\toprule
\rowcolor{Gray}
\textbf{Reference} & \textbf{Approach} & \textbf{Cancer types} & \textbf{\#Sample} & \textbf{Data type} & \textbf{Acc} \\\midrule
    Karim et al.~\cite{karim2018a2ic} & DBN/LSTM & 14 primary types & 15,699 & TCGA CNVs & 73\% \\
    Yuan et al.~\cite{yuan2018cancer} & CNN & 25 primary types & 14,703 & CNA/cromatin & 90\% \\% 2018
    Elsadek et al.~\cite{elsadek2018supervised} & LR & 6 primary types & 3,480  & CNV & 85\% \\% 2018
	Cruz et al.~\cite{19Cruz} & CNN & Breast cancer  & 605 & Slide image & 96\%  \\% 2017
    Danee et al.~\cite{17Danaee} & MLP & Breast cancer & 1210 & RNA-seq & 94\% \\% 2016
    Ning et al.~\cite{zhang2016classification} & Dagging & 6 primary types & 3,480  & CNV & 75\% \\% 2016
	Rajana et al.~\cite{20Rajanna} & Deep NN & Prostate cancer & 807 & Histology & 95\%\\
    Chen et al.~\cite{18Chen} & Shallow NN & Colon cancer  & 590 & Gene-exp & 84\% \\% 2015
   % Zheng et al.~\cite{23Zheng} & K-means/SVM & Breast cancer & 569 & Phenotype & 97\% \\% 2014
    %Xiaofan et al.~\cite{ding2014application} & NaÃ¯ve Bayes & Cancer risk & 640 & Human SNP & 93\% \\ % 2014
    \bottomrule
\end{tabular}
\end{center}
\end{table*}
\fi 
Since DL algorithms work better with high dimensional data, recent studies focused on employing NN architectures, which in comparison with traditional ML-based approaches, have shown more accurate and promising results for cancer identification. In particular, CNN has shown tremendous success and has gained much attention for solving gene selection and classification based on microarray data~\cite{zeebaree2018gene}. Further, Cruz et al.~\cite{19Cruz} trained a CNN using whole slide images, and extract deep features from different cohorts, which are used to detect cancer regions. % with a very high degree of precision. 
%Since DL algorithms can work better with high dimensional data; recent studies focused on employing deep neural network~(DNN) architectures, particularly for breast cancer. 
Danaee et al.~\cite{17Danaee} used a stacked denoising autoencoder to extract features from RNA-seq data, which are then fed into a SVM and a shallow ANN to classify malignant or benign tumors of breasts~\cite{18Chen}.
%DeepCNA is another CNN-based approach proposed for cancer type prediction based on CNVs and chromatin 3D structure with CNN~\cite{yuan2018cancer}. 
%Other approaches used CNVs for cancer risk and type predictions~\cite{ding2014application, zhang2016classification, elsadek2018supervised}. E.g., Ding et al.~\cite{ding2014application} used recurrent CNVs from non-tumor blood cell DNA of non-cancer subjects about hepatocellular carcinoma, gastric cancer, and colorectal cancer patients. %They found to reveal the differences between cancer patients and controls concerning CN losses and CN gains.
Although their study makes predictions on the cancer predisposition of unseen test groups of mixed DNAs with high confidence, it is limited to only Caucasian and Korean cohorts. 
Elsadek et al.~\cite{elsadek2018supervised} used CNVs about 23,082 genes for 2,916 instances from cBioPortal to classify the tumor samples of breast, bladder urothelial, colon, glioblastoma, kidney, and head and neck squamous cells and achieve an accuracy of 85\%. %Their study indicates that only a few genes may play important roles in differentiating cancer types. %Then, Elsadek et al.~\cite{elsadek2018supervised} used the same dataset and extended it by training 7 ML classifiers in which the random forest algorithm shows 86\% accuracy. %The data used the CNV invariant types of cancers were downloaded from the 

Lyu et al.~\cite{lyu2018deep} and Mostavi et al.~\cite{mostavi2019convolutional} embedded the RNA-Seq data from the PCA project into 2D images and trained a CNN to classify 33 tumor types, which outperforms the approach in~\cite{li2017comprehensive}.
%on the same dataset. 
Besides, they provide a functional analysis on the genes with high intensities in the HM based on GradCAM and validated that these top genes are related to tumor-specific pathways. However, due to the stochastic nature of NN, the prediction and feature importance generated is slightly different across runs, i.e., not deterministic. This is also no exception for tree-based ensemble models such as gradient boosted trees~(GBT), which provides 3 options for measuring feature importance: i) \emph{weight}, which is the number of times a feature is used to split the data across all trees, ii) \emph{cover}, the number of times a feature is used to split the data across all trees weighted by the number of training data points go through those splits, and iii) \emph{gain}, which is the average training loss reduction gained when using a feature for splitting. Based on these measure, feature importance orderings~(i.e., the order in which features were added) are different since subsequent features will get a disproportionately high weight.

Our proposed approach \emph{OncoNetExplainer} first embeds high dimensional RNA-Seq data into 2D images and trains CNN and VGG16 networks with GradCAM++ activated to classify 33 tumor types based on patients' GE profiles and provides a human-interpretable explanation~(post-model explainability) to identify important biomarkers, which is further validated based on the annotations from TumorPortal. To provide a comparison with baselines, we also identify the top-K biomarkers for each cancer type and cancer specific driver genes 
%across 33 cancer types 
based on GBT and SHAP~(pre-model explainability). 
%The final accuracy we got was 95.59\%, higher than another paper applying the GA/KNN method on the same dataset. Based on the idea of Guided Grad Cam, as to each class, we generated significance heat map for all the genes. By doing functional analysis on the genes with high intensities in the heat maps, we validated that these top genes are related to tumor-specific pathways, and some of them have already been used as biomarkers, which proved the effectiveness of our method. As far as we know, we are the first to apply convolutional neural network on Pan-Cancer Atlas for classification, and we are also the first to match the significance of classification with the importance of genes. Our experiment results show that our method has a good performance and could also apply in other genomics data.

\section{Materials and Methods}
\label{dc}
In this section, we discuss the data preparation, network construction, training, and biomarkers discovery with ranking.
\vspace{-4mm} 
\subsection{Data collection and preparation}
We use the cancer transcriptomes from the Pan-Cancer Atlas project to interrogate GE states induced by deleterious mutations and copy number alterations. In particular, GE profiles about 33 prevalent tumor type for 9,074 samples are used in our approach. This dataset has been used widely as prior knowledge to generate tumor-specific biomarkers~\cite{way2018machine,hoadley2018cell,malta2018machine}. These data are hybridized by the Affymetrix 
%Genome-Wide Human SNP Array 
6.0
%\footnote{\url{http://www.affymetrix.com/support/technical/byproduct.affx}}
, which allows us to examine the largest number of cases along with the highest probe density~\cite{31Park}. \Cref{table:datadetails} shows sample distribution. 
\begin{table} [!ht] 
\begin{center}
\scriptsize
\caption{Sample distribution across different tumor types}
\vspace{-1mm} 
\label{table:datadetails}
\begin{tabular}{lll}
\toprule
%\rowcolor{Gray}
 \textbf{Cohort} & \textbf{\#Sample} & \textbf{Carcinoma type} \\\midrule
 BRCA & 981  & Breast invasive carcinoma  \\%
 LGG  & 507  & Brain lower grade glioma \\% 
 UCEC & 507  & Uterine endometrial carcinoma \\%
 LUAD & 502  & Lung adeno-carcinoma \\%
 HNSC & 487  & Head-neck squamous cell carcinoma \\% 
 THCA & 480  & Thyroid carcinoma \\%
 PRAD & 479  & Prostate adeno-carcinoma \\%
 LUSC & 464  & Lung squamous cell carcinoma \\%
 BLCA & 398  & Bladder urothelial carcinoma \\%
 STAD & 383  & Stomach adeno-carcinoma \\%
 SKCM & 363  & Skin cutaneous melanoma \\%
 KIRC & 352  & Kidney renal clear cellcarcinoma  \\%
 LIHC & 348  & Liver hepato-cellular carcinoma	\\%
 COAD & 341  & Colon adeno-carcinoma \\%
 CESC & 272  & Cervical \& endo-cervical cancer \\%
 KIRP & 271  & Kidney papillary cell carcinoma	\\%
 SARC & 229  & Sarcoma \\%
 OV   & 176  & Ovarian serouscystadenocarcinoma \\%
 ESCA & 169  & Esophageal carcinoma \\%
 PCPG & 161  & Pheochromocytoma-paraganglioma \\%
 PAAD & 152  & Pancreatic adenocarcinoma	\\%
 TGCT & 144  & Testicular germ cell tumor \\%
 GBM  & 124  & Glioblastoma multiforme \\%
 THYM & 119  & Thymoma	 \\%
 READ & 118  & Rectum adeno-carcinoma \\%
 LAML & 115  & Acute myeloid leukemia	\\%
 MESO & 82  & Mesothelioma	\\%
 UVM & 80  & Uveal melanoma	 \\%
 ACC & 76  & Adrenocortical cancer	\\%
 KICH & 65  & Kidney chromophobe	\\%
 UCS & 56  & Uterine carcino-sarcoma	 \\%
 DLBC & 37  & Diffuse large B-cell lymphoma	\\%
 CHOL & 36  & Cholangio carcinoma	 \\%
 \bottomrule
\end{tabular}
\vspace{-3mm} 
\end{center}
\end{table}
To apply the convolutional~(conv) operations, 
%for finding important biomarkers using GradCAM and MAI
we embed GE samples into 2D images in which GE values for each sample are reshaped from a 20,501$\times$1 array into a 144$\times$144 image by zero padding around the edges and normalized to [0,255]
%. %All the images are then normalized to [0,255].% 
without losing any information. 

%\vspace{-2mm} 

\begin{figure*}
	\vspace{-2mm} 
	\centering
	\includegraphics[width=0.85\textwidth,height=53mm]{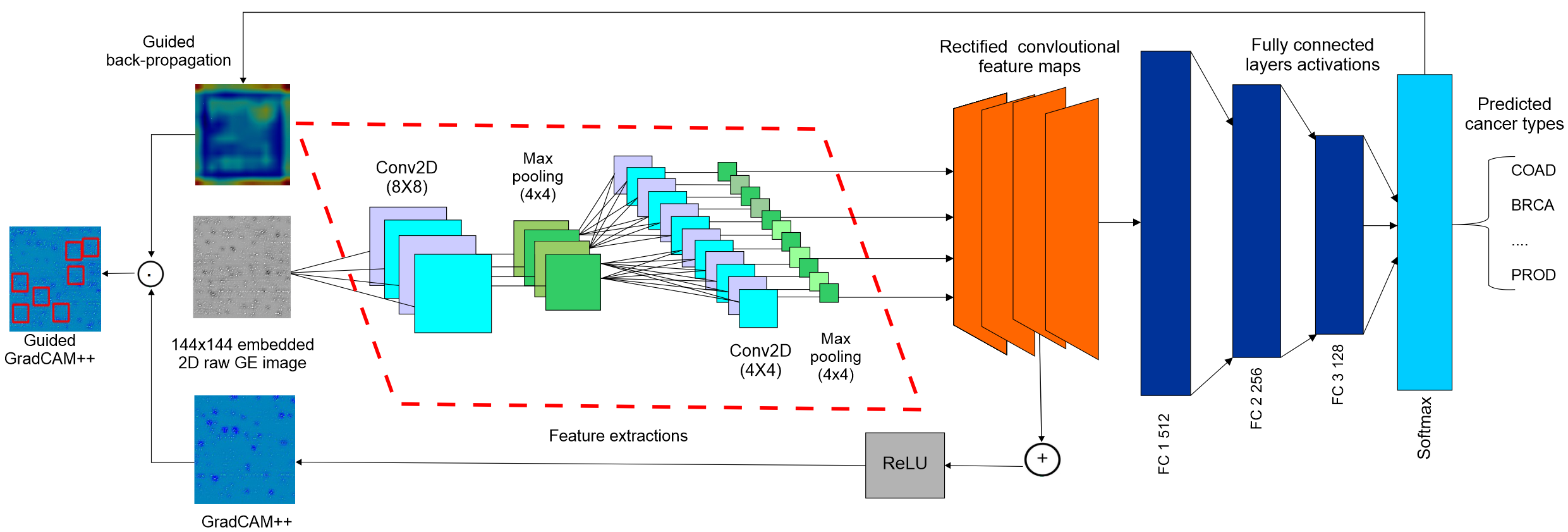}	
    \caption{Schematic representation of our approach, which starts from taking a raw GE sample and passing to conv layers before obtaining rectified conv feature maps~(with guided-backprop \& GradCAM++) to pass through dense, dropout, \& softmax layer}	
	\label{fig:clstm}
	\vspace{-5mm} 
\end{figure*}

\begin{algorithm*}
\DontPrintSemicolon \SetKwInOut{Input}{Input}%
  \SetKwInOut{Output}{Output}%
  \Input{2D GE images $\mathcal{D}=(d_1,d_2,\dots,d_n)$ having ground truth~(i.e., labels) $\mathcal{L}=(l_1,l_2,\dots,l_j)$ on which a CNN model is trained for each fold $\mathcal{M}=(m_1,m_2,\dots,m_i)$ to find $k$ for top-k genes that satisfy MAI threshold.}%
  \Output{feature importance $\mathcal{F}=(f_1,i_1)(f_2,i_2),\dots,(f_n, i_n)$ and top features $\mathcal{T}$ across all images per fold per class.}%
  \BlankLine%
 \For{$\mathit{fold} \in \mathit{FOLDS}$}{
   $\mathcal{P} \leftarrow  \{\}$ \tcp*{Guided backprop for each image per fold per class}\; 
  \vspace{-4mm} 
   $\mathcal{K} \leftarrow  \{\}$ \tcp*{GradCAM++(GCAM) for all images per fold per class}\; 
  \vspace{-4mm} 
   $\mathcal{I} \leftarrow  \{\}$ \tcp*{GCAM of each image in a fold}\; 
  \vspace{-4mm} 
  $\mathcal{G} \leftarrow  \{\}$ \tcp*{GCAM for all images per fold per class}\; 
  \vspace{-4mm} 
  $\mathcal{F} \leftarrow  \{\}$ \tcp*{Feature importance of each gene per class per fold}\; 
  \vspace{-4mm} 
  $\mathcal{T} \leftarrow  \{\}$ \tcp*{Top genes and importance per class per fold}\; 
  	\vspace{-4mm} 
  	\For {$d \in \mathcal{D}$}{ 
		$\mathcal{K} \leftarrow \mathit{gradCAM++}(m_{d}, d, l_d)$ \tcp*{GCAM of images per fold per class}\;
		\vspace{-4mm} 
		$\mathcal{P} \leftarrow \mathit{guidedBackprop}(m_{d}, d)$ \tcp*{Guided backprop of each image}\;
		\vspace{-4mm} 
		$\mathcal{I} \leftarrow \mathcal{K} * \mathcal{P}$ \tcp*{GCAM of each image}\;
		\vspace{-4mm} 
		$\mathcal{G} \leftarrow \mathcal{G}\cup \mathcal{I}$ \tcp*{GCAM for all the images in fold}\;
		\vspace{-4mm} 
	    } 
	$\mathcal{F} \leftarrow \frac{1}{N}\sum_{i=1}^N \mathcal{G}$ \tcp*{Mean absolute impact for genes for axis=0}\;
	\vspace{-4mm} 
	\If{$F_i < \sigma $}{ \tcp*{If the feature importance is less than MAI}\;
	\vspace{-4mm} 
    	$\mathcal{F} \leftarrow  F - F_i$ \tcp*{Pop off insignificant genes}\;
    	\vspace{-4mm} 
    	$\mathcal{T} \leftarrow  \operatorname{sort}_k(F)$ \tcp*{Sort and choose top genes based on MAI}\;
    	\vspace{-4mm} 
	}
	\textbf{Return} $\mathcal{F}, \mathcal{T}$
 }
 \caption{Computing feature importance and ranking genes}
 \label{algo:hm}
\end{algorithm*}

\begin{algorithm*}
\DontPrintSemicolon \SetKwInOut{Input}{Input}%
  \SetKwInOut{Output}{Output}%
  \Input{importance of current class across folds $\mathcal{F}=(f_1,\dots,f_i)$, height $h$ \& width $w$ of rectangle, \& MAI threshold $\sigma$.}%
  \Output{important areas $\mathcal{C} = (x_{1}, y_{1}),(x_{2}, y_{2}),\dots,(x_{n}, y_{n})$ in an image per fold.}%
  \BlankLine%
 \For{$\mathit{fold} \in \mathit{FOLDS}$}{
  $\mathcal{A} \leftarrow \mathit{dict}()$ \tcp*{Importance of areas}\; 
  \vspace{-4mm} 
  $\mathcal{S} \leftarrow \mathit{list}()$ \tcp*{Sorted areas by MAI}\; 
  \vspace{-4mm} 
  $\mathcal{C} \leftarrow \mathit{list}()$ \tcp*{Important areas}\; 
  \vspace{-4mm} 
  \For {$h$}{ 
    	\For {$w$}{ 
		$area \leftarrow \mathcal{F}[h:h + \mathit{shape}[0], w:w + \mathit{shape}[1]]$ \tcp*{Area of image}\;
		\vspace{-4mm} 
		$impA \leftarrow 0$\tcp*{Importance of current area in the image}\;
		\vspace{-4mm} 
		\For {$\mathit{row} \in \mathit{area}$}{
		%\vspace{-4mm} 
			\For {$\mathit{imp} \in \mathit{row}$}{
			%\vspace{-4mm} 
			\If{$\mathit{imp} > \sigma$}{ \tcp*{If feature importance is greater than MAI}\;
			\vspace{-4mm} 
			    $\mathit{impA} += \mathit{imp} - \sigma$ \tcp*{Importance of area = current importance - MAI}\;
			    \vspace{-4mm}
			    }
			  }
		}
        $\mathcal{A}[\mathit{area}] = \mathit{impA}$ \tcp*{We update the importance of the area}
	     } 
	    }
	$\mathcal{S} \leftarrow \operatorname{sort}(\mathcal{A}, \mathit{reverse}=\mathit{true})$ \;
	\For {$a, i \in \mathcal{S}$}{
	\If{$a \cap i = \empty$}{ \tcp*{Non-intersecting area with important areas}\;
		\vspace{-4mm} 
    	$\mathcal{C} \leftarrow \mathcal{C}  \cup a$ \tcp*{It's a new important area added to the list}\;
    		\vspace{-4mm} 
	}}
	\textbf{Return} $\mathcal{C}$
 }
 \caption{Identification of important areas}
 \label{algo:imp_area}
\end{algorithm*}

\subsection{Network construction and training}
\label{nc}
We trained a shallow CNN from scratch alongside data augmentation in which the output of each conv layer is passed to dropout and Gaussian noise layers to avoid overfitting and thus regularize the learning~\cite{vardropout}. This involves the input feature space into a lower-dimensional representation, which is then further down-sampled by two different pooling layers and a max-pooling layer~(MPL) by setting the pool size. The output of an MPL is considered as an `extracted feature' from each 2D GE image. Since each MPL `flattens' the output space by taking the highest value in a FM, this produces a sequence vector from the last conv layer, which we expect to force the GE value of specific genes that are highly indicative of being responsible for a specific cancer type. Then this vector is passed through another dropout layer and a fully connected softmax for the probability distribution over the classes. 

The CNN is trained with %the first-order gradient-based optimization 
AdaGrad to optimize the categorical cross-entropy loss of the predicted cancer type vs.\ actual cancer type. Further, we observe the performance by adding the Gaussian noise layer~(GNL) following each conv layer to improve model generalization. 
Further, we used the pretrained VGG16 network to which we added two dense layers at the end of the original architecture, followed by a GNL. Then, we fine-tuned the top layers with minor weight updates: 

\vspace{1mm} 
\begin{itemize}
    \item First, we instantiated the conv base of the VGG16 network and loaded its weights.
    \item Then, we added our previously defined fully-connected layers on top with minor weight updates.
    \item  Finally, we placed a softmax layer by freezing up to the last conv block of the VGG16 model, which yields a probability distribution over 33 different classes. 
\end{itemize}
\vspace{1mm} 

Since the data is very high dimensional, we chose not to go for manual feature selection. Rather, we let both CNN and VGG16 networks extract the most important features. The guided back-propagation helps to generate more human-interpretable but fewer class-sensitive visualizations than the saliency maps~(SM)~\cite{nie2018theoretical}. Since SM use true gradients, the trained weights are likely to impose a stronger bias towards specific subsets of the input pixels.
Accordingly, class-relevant pixels are highlighted rather than producing random noise~\cite{nie2018theoretical}. Therefore, GradCAM++ is used to draw the HM to provide attention to most important genes.
Class-specific weights of each FM are collected from the final conv layer through globally averaged gradients~(GAG) of FMs instead of pooling~\cite{chattopadhay2018grad}: 

\vspace{-1mm} 
\begin{equation}
\alpha_k^c=\frac{1}{Z}\sum_{i}\sum_{j}\frac{\partial y^c}{\partial A_{ij}^k}
\label{eq:alpha}
\end{equation}
\vspace{-3mm} 

where $Z$ is the number of pixels in a FM, $c$ is the gradient of the class, and $A_{ij}^k$ is the value of $k$th FM at $(i,j)$. Having gathered relative weights, the coarse SM, $L^c$ is computed as the weighted sum of $\alpha_k^c*A_{ij}^k$ of the ReLU activation function and employ the linear combination to the FM, since only the features with positive influence on the class are of interest~\cite{chattopadhay2018grad}.% and the negative pixels that belong to other categories in the image are discarded~\cite{114}:

\vspace{-3mm} 
\begin{equation}
L^c=\operatorname{ReLU}(\sum_{i}\alpha_k^cA^k)
\label{3.11}
\end{equation}
\vspace{-3mm} 

The GradCAM++ replaces the GAG with a weighted average of the pixel-wise gradients (\cref{eq:w}), since the weights of pixels contribute to the final prediction(\cref{3.13}) by aggregating \cref{eq:w} and $\alpha_{ij}^{kc}$ (\cref{3.14}). In summary, it applies the following iterators over the same activation map $A^k$, $(i,j)$ and $(a,b)$:

\vspace{-3mm} 
\begin{equation}
w_k^c=\sum_{i}\sum_{j}\alpha_{ij}^{kc}\cdot \operatorname{ReLU}(\frac{\partial y^c}{\partial A_{ij}^k})
\label{eq:w}
\end{equation}
\vspace{-3mm} 
\begin{equation}
y^c=\sum_{k}w_k^c\cdot \sum_{i}\sum_{j}A_{ij}^k
\label{3.13}
\end{equation} 
\vspace{-2mm} 
\begin{equation}
\alpha_{ij}^{kc}=\frac{\frac{\partial^2y^c}{(\partial A_{ij}^k)^2}}{2\frac{\partial^2y^c}{(\partial A_{ij}^k)^2}+\sum_{a}\sum_{b}A_{ab}^k{\frac{\partial^3y^c}{\{(\partial A_{ij}^k)^3\}}}}
\label{3.14}
\end{equation}

Further, since an appropriate selection of hyperparameters can have a huge impact on the performance of a deep architecture, we perform the hyperparameter optimization through a random search and 5-fold cross-validation tests. In each of 5 runs, 70\% of the data is used for the training, 30\% for the evaluation.
10\% of the training set is used for validation of the networks to optimize the cross-entropy loss based on the best learning rate, batch size, number of filters, kernel size, and dropout/Gaussian noise probability. 

\subsection{Finding and validating important biomarkers}
\Cref{algo:hm} and \ref{algo:imp_area} depict the pseudocodes for computing feature importance with ranking genes and identification of important areas on the HM, respectively. %After executing these, 
We averaged all the normalized HM from the same class to generate a class-specific HM inspired by Selvaraju et al.~\cite{chattopadhay2018grad}. In the HM, a higher intensity pixel represents a higher significance to the final prediction, which indicates higher importance of corresponding genes and the GE values. Top genes are then selected based on the intensity rankings and MAI threshold. Since GradCAM++ requires all the samples to run through the network once, we let the trained CNN models set and record the activation maps in the forward pass, and the gradient maps in the back-prop to collect the HM for each sample.% from the trained model. %The idea is also to find the most important biomarkers by ranking them based on MAI threshold.

In contrast, Shapley values are used to calculate the importance of a feature by comparing what a model predicts with and without a feature from all possible combinations of $n$ features in the dataset $S$. Given a GE value of feature $i \in S$, SHAP calculates the prediction $p$ of the model with $i$. The Shapely value $\phi$ is calculated as follows~\cite{NIPS2017_7062}: 

\vspace{-4mm} 
\begin{align}
    \phi_{i}(p)=\sum_{S \subseteq N / i} \frac{|S| !(n-|S|-1) !}{n !}(p(S \cup i)-p(S))
    \label{eq:shap}
\end{align}
\vspace{-3mm} 

However, since the order in which a model sees features can affect the predictions, this computation is repeated in all possible orders %\footnote{P(2050, 2) = 2050!/(2050−2)! = 4,200,450 possible orders} 
to compare the features fairly. Feature that have no effect on the predicted value are expected to produce a Shapley value of 0. However, if two features contribute equally to the prediction, the Shapley values should be the same~\cite{NIPS2017_7062}. 

\section{Experiments}
\label{results}
Implementation was done in Python\footnote{\url{https://github.com/rezacsedu/XAI_Cancer_Prediction}} using % and experiments were carried out on a machine having an Intel(R) Xeon(R) CPU E5-2640, 256 of RAM, and Ubuntu 16.04 OS.
a software stack comprising Scikit-learn and Keras with the TensorFlow backend.
The network was trained on an Nvidia GTX 1080i GPU with CUDA and cuDNN enabled. % to make the overall pipeline faster.
%For the experiment, 80\% of the data is used for the training and evaluate the optimized model on 20\% held-out data in which the best hyperparameters were .
Results based on hyperparameters produced through random search and 5-fold cross-validation are reported and discussed with a comparative analysis with macro-averaged precision and recall. Further, since the classes are imbalanced, Matthias correlation coefficient~(MCC) scores were reported. 
%We did not report F1-scores since it is significant only when the value of precision and recall are very different. 
Since it is important for cancer diagnosis to have both high precision and high recall~\cite{naulaerts2017precision}, results with very different precision and recall are not useful in cancer diagnosing and tumor type identification. Hence, we did not report F1-scores.  

\subsection{Performance of cancer type classification}
The average accuracy obtained was 89.75\% and 96.25\% using CNN and VGG16 models, respectively. However, since the classes are imbalanced, only the accuracy will give a very distorted estimation of the cancer types. Thus, we report the class-specific classification reports along with the corresponding MCC scores in \cref{table:class_specific}. 
As can be seen, precision and recall for the majority cancer types were high and for these the VGG16 model performs mostly better. 
Notably, the VGG16 model classifies BRCA, UCEC, LUAD, HNSC, LUSC, THCA, PRAD, BLCA, STAD, KIRC, LIHC, COAD, CESC, KIRP, SARC, OV, PCPG, TGCT, GBM, READ, LAML, MESO, and DLBC cancer cases more confidently, whereas the CNN model classifies PAAD, CHOL, and UCS cancer cases more accurately. 

\begin{table}[h]
\caption{Cancer type prediction: CNN vs VGG16}
\vspace{-3mm} 
\label{table:class_specific} %RF Confusion Matrix Oncogene Subtype
\begin{center}
\scriptsize
\begin{tabular}{l|lll|lll}
\toprule
%\rowcolor{Gray}
{} & \multicolumn{2}{c}{\textbf{\hspace{0.7cm} CNN~(89.75\%)}} & \multicolumn{3}{c}{ \textbf{\hspace{1.5cm} VGG16~(96.25\%)}} &  {} \\
\textbf{Type } & \textbf{Precision} &  \textbf{Recall}  & \textbf{MCC} & \textbf{Precision} &  \textbf{Recall} & \textbf{MCC} \\\midrule%\hline
BRCA   & {\color{red}\textbf{0.8785}} & 0.8612 & 0.7564 & {\color{red}\textbf{0.9437}} & 0.9511 & 0.8465  \\%\hline
LGG    & 0.9254 & 0.8926 & 0.8330 & 0.9311 & 0.9402 & 0.8421  \\%\hline
UCEC   & 0.8753 & 0.8819 & 0.7835 & 0.9562 & 0.9429 & 0.8445  \\%\hline
LUAD   & 0.8235 & 0.8354 & 0.7136 & 0.9865 & 0.9823 & 0.8624  \\%\hline
HNSC   & 0.8520 & 0.8743 & 0.7851 & 0.9730 & 0.9822 & 0.8765  \\%\hline
THCA   & {\color{red}\textbf{0.8528}} & 0.8323 & 0.7275 & {\color{red}\textbf{0.9138}} & 0.9154 & 0.8125  \\%\hline
PRAD   & {\color{red}\textbf{0.8827}} & 0.8778 & 0.7847 & {\color{red}\textbf{0.9233}} & 0.9347 & 0.8207  \\%\hline
LUSC   & 0.8726 & 0.8634 & 0.7625 & 0.9434 & 0.9472 & 0.8524  \\%\hline
BLCA   & 0.8956 & 0.9037 & 0.8075 & 0.9656 & 0.9537 & 0.8475  \\%\hline
STAD   & 0.8253 & 0.8156 & 0.6932 & 0.9653 & 0.9556 & 0.8532  \\%\hline
SKCM   & 0.8853 & 0.8711 & 0.8025 & 0.9046 & 0.9136 & 0.8168  \\%\hline
KIRC   & 0.8967 & 0.9123 & 0.8237 & 0.9578 & 0.9689 & 0.8531  \\%\hline
LIHC   & 0.8194 & 0.8085 & 0.6945 & 0.9572 & 0.9664 & 0.8537  \\%\hline
COAD   & 0.8368 & 0.8245 & 0.7679 & 0.9776 & 0.9690 & 0.8514  \\%\hline
CESC   & 0.8785 & 0.8743 & 0.7964 & 0.9873 & 0.9885 & 0.8664  \\%\hline
KIRP   & 0.8254 & 0.8032 & 0.7043 & 0.9681 & 0.9782 & 0.8430  \\%\hline
SARC   & 0.8753 & 0.8671 & 0.7835 & 0.9365 & 0.9435 & 0.8421 \\%\hline
OV     & 0.8825 & 0.8733 & 0.7936 & 0.9725 & 0.9773 & 0.8262  \\%\hline
ESCA   & {\color{cyan}\textbf{0.8913}} & 0.8719 & 0.7951 &  {\color{cyan}\textbf{0.8956}} & 0.8834 & 0.8076  \\%\hline
PCPG   & 0.8537 & 0.8611 & 0.7875 & 0.9875 & 0.9987 & 0.8735  \\%\hline
PAAD   & 0.9629 & 0.9567 & 0.8407 & 0.9452 & 0.9500 & 0.8325  \\%\hline
TGCT   & 0.8736 & 0.8722 & 0.7825 & 0.9890 & 0.9724 & 0.8434  \\%\hline
GBM    & 0.8952 & 0.8845 & 0.8075 & 0.9362 & 0.9453 & 0.8436  \\%\hline
THYM   & 0.9255 & 0.9123 & 0.8232 & 0.9775 & 0.9678 & 0.8622  \\%\hline
READ   & {\color{cyan}\textbf{0.6795}} & 0.6857 & 0.6225 & {\color{cyan}\textbf{0.8874}} & 0.8733 & 0.7525  \\%\hline
LAML   & 0.8697 & 0.8567 & 0.8237 & 0.9576 & 0.9632 & 0.8513  \\%\hline
MESO   & 0.8991 & 0.9028 & 0.8076 & 0.9534 & 0.9456 & 0.8457  \\%\hline
UVM    & 0.8765 & 0.8623 & 0.7979 & 0.9136 & 0.9089 & 0.8184  \\%\hline
ACC    & 0.9217 & 0.9345 & 0.8225 & 0.9623 & 0.9731 & 0.8611  \\%\hline
KICH   & 0.9335 & 0.9475 & 0.8425 & 0.9690 & 0.9625 & 0.8439  \\%\hline
UCS    & {\color{cyan}\textbf{0.9157}} & 0.9064 & 0.8125 & {\color{cyan}\textbf{0.8726}} & 0.8675 & 0.7869  \\%\hline
DLBC   & 0.8678 & 0.8729 & 0.7005 & 0.9347 & 0.9421 & 0.8389  \\%\hline
CHOL   & {\color{cyan}\textbf{0.8838}} & 0.8975 & 0.7979 & {\color{cyan}\textbf{0.8455}} & 0.8342 & 0.6821  \\%\hline
\midrule
%\rowcolor{LightCyan}
\textbf{Average} &   \textbf{0.8975}    &  \textbf{0.9065} &    \textbf{0.8052}   &  \textbf{0.9625} & \textbf{0.9542} & \textbf{0.8453}\\
\bottomrule
\end{tabular}
\end{center}
\vspace{-8mm} 
\end{table}

\begin{figure}[h]
  	\vspace{-3mm} 
	\centering
		\centering
		\includegraphics[width=0.98\linewidth,height=63mm]{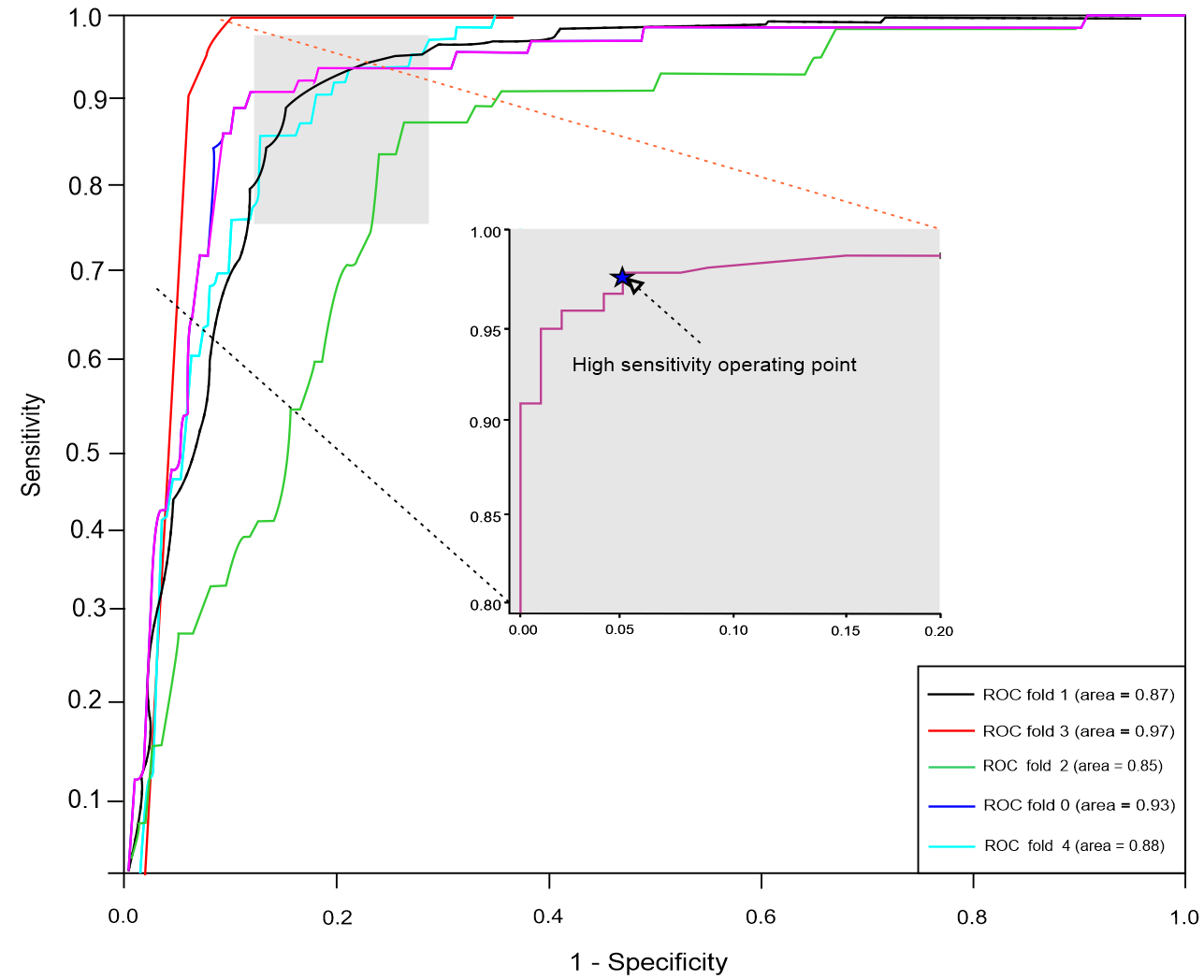}
  	\caption{ROC curves of the VGG16 model across folds} 
  	\vspace{-1mm} 
	\label{roc:both_dataset}
\end{figure}

%\subsection{Validation of top genes}
\begin{figure*}[h]
	\vspace{-2mm} 
		\centering
		\includegraphics[width=0.8\linewidth,height=76mm]{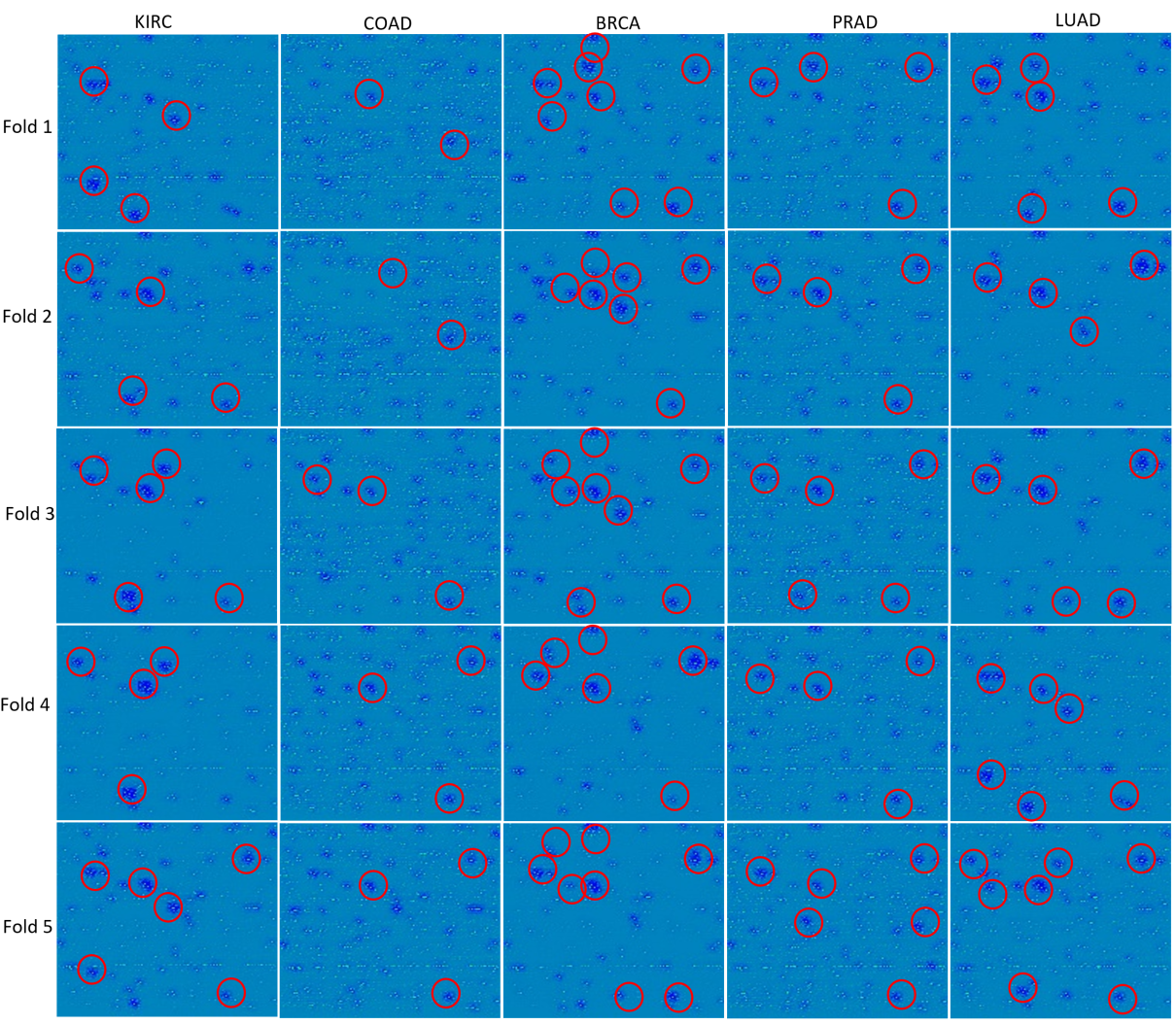}
	\caption{Heat map examples for selected cancer types. Each row represents the result from one fold. Columns represent the heat maps of BRCA, KIRC, COAD, LUAD, and PRAD cancer types (patterns are not clearly visible in some folds, though)} 
	\label{fig:hm}
		\vspace{-4mm} 
\end{figure*}

\begin{figure}
    \centering
		\includegraphics[width=\linewidth,height=65mm]{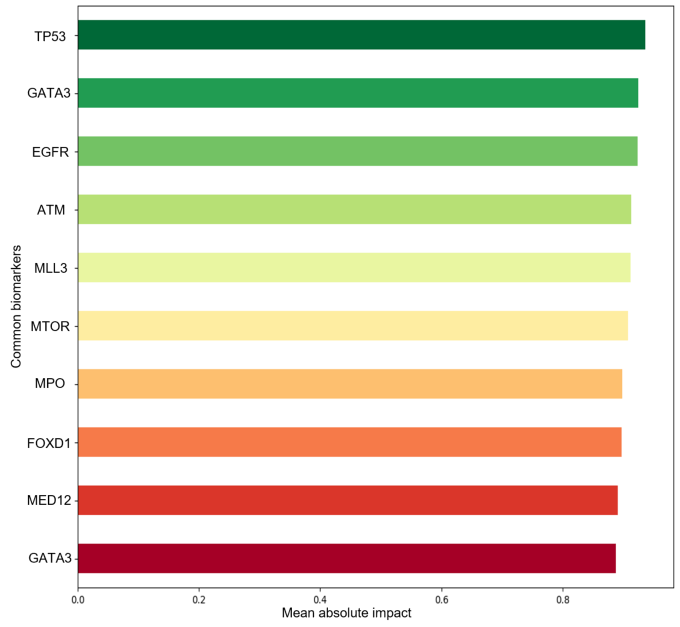}
	    \caption{Common driver genes across 33 cancer types)} 
    	\label{fig:commongenes}
    	\vspace{-5mm} 
\end{figure}

%The downside is that both classifiers made substantial mistakes, e.g., VGG16 could classify HNSC and LUSC tumor samples accurately in only 79\% and 81\% of the cases. On the other hand, the CNN model made more mistakes particularly on the STAD, HNSC, LUSC, and LGG tumor samples. 
%In summary, both classifiers performed moderately well except for certain types of tumor cases such as STAD, HNSC, BLCA, THCA, UCEC, LUAD, LUSC, and LGG. 

The ROC curves generated by the VGG16 model in \cref{roc:both_dataset} show that the AUC scores are consistent across the folds.% showing stable predictions.%, which shows about 5\% boost in AUC scores generated by the VGG16 model. 
This signifies that the predictions by the VGG16 model are much better than random guessing. Further, the class-specific MCC
%\footnote{Measured using a Pearson product-moment correlation coefficient}
scores of the VGG16 model is 4\% higher than that of the CNN model, which suggests that the predictions were strongly correlated with the ground truth, yielding a Pearson product-moment correlation coefficient higher than 0.70 for all the classes except for the CHOL tumor samples. 
The downside, however, is that both classifiers made a number of mistakes too, e.g., VGG16 can classify ESCA, READ, UCS, and CHOL tumor cases in only 89\% of the cases accurately, while the CNN model made more mistakes particularly for the READ, LUAD, LIHC, KIRP, COAD, and STAD tumor samples. 

\subsection{Feature importance and validation of top biomarkers}
We identified top genes for which the change in expression has significant impact on patients. \Cref{fig:hm} shows examples of HM generated for each class
%~(each column for one cancer type)
across 5 different folds. As seen, there are similarities across folds and displaying distinct and similar patterns when comparing different cancer types. The red circles highlight similar patterns, e.g., between KIRC and BRCA, and PRAD and LUAD across folds, whereas COAD shows very different patterns. Although there are differences among folds, some patterns are clearly visible. 
Since intensities did not follow any regular pattern, we chose top 660 genes across 33 tumor types~(top-20 genes per class) as more significant based on the measure of MAI. Since we have more than 20K protein-coding genes, our choice of 660 is still a reasonable choice, since the number of important biomarkers should be small whose GE changes are sensitive to cancer growth~\cite{zuo2019identification}.
All genes in the top-20 list can thus be viewed as tumor-specific biomarkers, which contribute most toward making the predictions. As for the other 29 tumor types, only 3 genes were in the list.
%and are not reported. 
Further hyperparameter tuning and training of both CNN models might improve this outcome. 

Then we further narrowed down the list to the top-5 genes in which only 5 tumor types~(i.e., BRCA, KIRC, COAD, LUAD, and PRAD) have at least five genes with feature importance of at least 0.5 w.r.t. MAI;
%. 
%Top-5 biomarkers with the MAI higher than 0.5 are 
they are shown in \cref{table:proteinimportance}.% from the 20,500 protein-coding genes.
To further validate our findings, the saturation analyses of cancer genes across 33 tumor types~(except for COAD) are obtained from the TumorPortal
%\url{http://www.tumorportal.org/}
\cite{lawrence2014discovery}. Validation for the COAD cancer follows a signature-based approach~\cite{zuo2019identification}, which was used for predicting the progression of colorectal cancer. However, our approach makes some false identifications, as 21 out of 25 genes are validated to be correct, making only 4 false identifications. 

\begin{table}
\caption{Top-5 genes and their importance}
\vspace{-3mm} 
\label{table:proteinimportance} %RF Confusion Matrix Oncogene Subtype
\begin{center}
\scriptsize
\begin{tabular}{llll}
\toprule
\textbf{Type} & \textbf{Gene} & \textbf{Gene type} & \textbf{MAI} \\ \midrule
\multirow{5}{*}{BRCA} & TP53 & Oncogene & 0.78125 \\ %\cline{2-4}
& GATA3 & Protein-coding & 0.760784 \\ %\cline{2-4}
& MLL3  & Protein-coding & 0.664706 \\ %\cline{2-4}
& TBX3  & Oncogene & 0.574118 \\ %\cline{2-4}
& MPO   & Protein-coding & 0.538039 \\ \midrule
\multirow{5}{*}{KIRC} & MTOR & Oncogene & 0.596078 \\ %\cline{2-4}
& SETD2 & Protein-coding  & 0.560784 \\ %\cline{2-4}
& ATM & Protein-coding & 0.540784 \\ %\cline{2-4}
& MPO & Oncogene & 0.531569 \\ %\cline{2-4}
& AMBN & Oncogene & 0.523137 \\ %\cline{2-4}
\midrule
\multirow{5}{*}{LUAD}& EGFR & Oncogene & 0.860000 \\ %\cline{2-4}
& KEAP1 & Protein-coding & 0.820784 \\ %\cline{2-4}
& ERBB2 & Oncogene & 0.764706 \\ %\cline{2-4}
& MLL3 & Protein-coding & 0.674118 \\ %\cline{2-4}
& AMBN & Protein-coding & 0.558039 \\ %\cline{2-4}
\midrule
\multirow{5}{*}{PRAD}& FOXA1 & Oncogene & 0.556078 \\ %\cline{2-4}
& TP53 & Oncogene & 0.520784 \\ %\cline{2-4}
& ATM & Protein-coding & 0.510784 \\ %\cline{2-4}
& AMBN & Protein-coding & 0.491569 \\ %\cline{2-4}
& MED12 & Protein-coding & 0.453137 \\ %\cline{2-4}
\midrule
\multirow{5}{*}{COAD}& EPHA6 & Protein-coding & 0.756078 \\ %\cline{2-4}
& TIMP1 & Protein-coding & 0.720784 \\ %\cline{2-4}
& ART5 & Protein-coding & 0.680784 \\ %\cline{2-4}
& FOXD1 & Protein-coding & 0.661569 \\ %\cline{2-4}
& AMBN & Protein-coding & 0.563137 \\ %\cline{2-4}
\bottomrule
\end{tabular}
\vspace{-8mm} 
\end{center}
\end{table}

\begin{figure*}[h]
\vspace{-3mm}
\centering
	\includegraphics[width=0.7\linewidth,height=17mm]{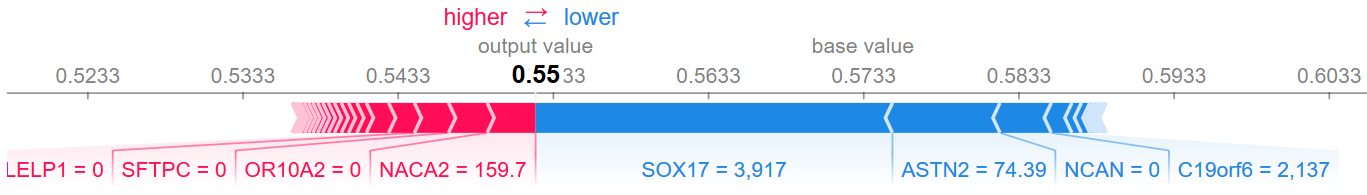}
	\caption{Clinical features' contribution for the first prediction: pushing the prediction higher and lower in red and blue, respectively} 
	\label{fig:shap}
\vspace{-4mm}
\end{figure*}

\subsection{Finding common biomarkers}
Identifying all significant common genes will help understand various aspects for a specific cancer type~(e.g., BRCA carcinogenesis). Thus, these top genes have close relations to the corresponding tumor types, which could be viewed as potential biomarkers. 
\Cref{fig:commongenes} shows the top-10 common biomarkers,
%across 33 cancer types. 
in which KRTAP1-1, INPP5K, GAS8, MC1R, POLR2A, BET1P1, NAT2, PSD3, KAT6A, and INTS10 genes are common across cancer types, with the INTS protein-coding gene having the highest feature importance of 0.6. 

\begin{figure}
\vspace{-5mm}
\centering
	\includegraphics[width=0.80\linewidth,height=85mm]{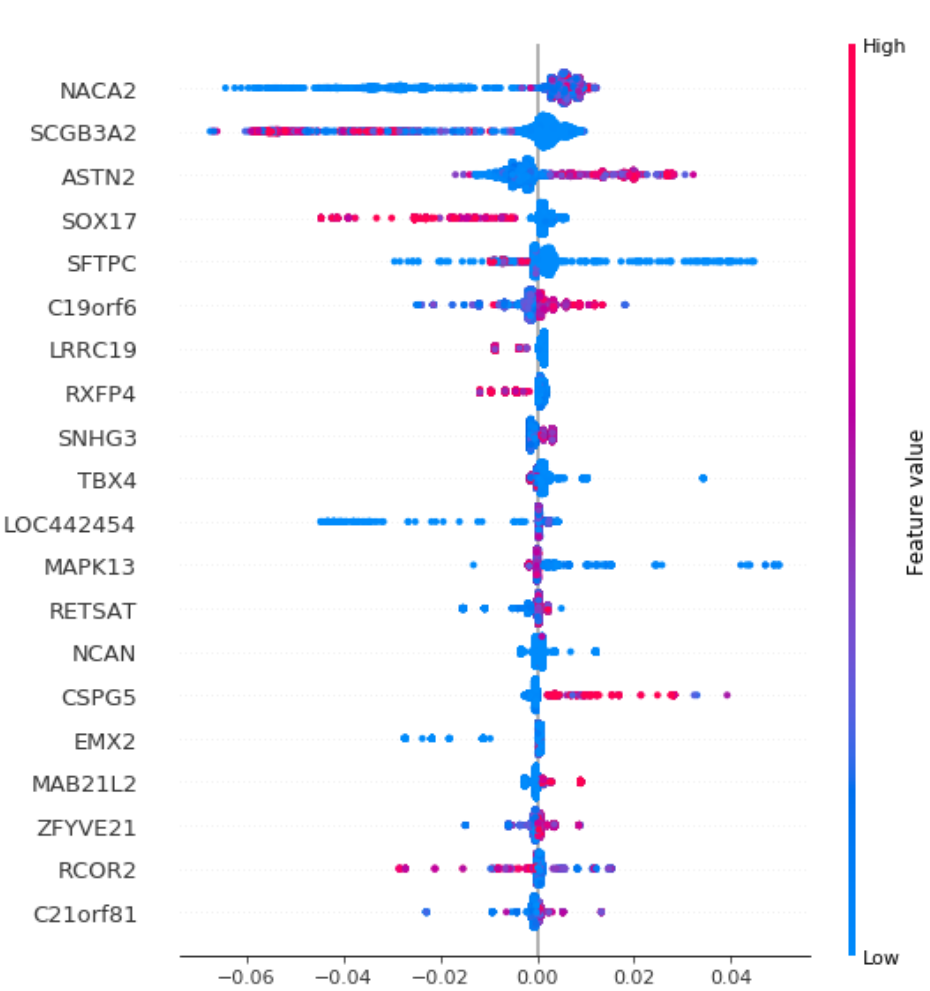}
	\caption{Clinical features ordered by ascending importance on the y-axis~(dots represent SHAP values of specific features)} 
	\label{fig:shap_FI}
\vspace{-5mm}
\end{figure}

\subsection{Explanations with SHAP}
The GBT model is trained to provide explanations generated by SHAP. \Cref{fig:shap} shows a base value that indicates the direction of the first prediction made by the GBT model and shows how much each feature is pushing the model's output from the base value\footnote{The average model output over the training dataset passed} 0.55 to the predicted output. Features pushing the prediction higher are shown in red; those pushing the prediction lower are in blue.
Further, to get an overview of which biomarkers are most important for the GBT model, we plot the SHAP values of each feature for each sample. The plot in \cref{fig:shap_FI} sorts features by the sum of SHAP value magnitudes over all the samples, shows the distribution of the impact of each feature on the model output, and gives the top-20 common biomarkers, where red represents high feature values, blue low. This reveals, e.g., that a low NACA2~(low GE value) lowers the predicted value. 
%across 33 cancer types 
Since the common biomarkers predicted by VGG16~(\cref{fig:commongenes}) and GBT~(\cref{fig:shap_FI}) are very different, a more detailed analysis of biological signaling pathways is further required to validate these findings. 

\subsection{Comparison with related works}
\emph{OncoNetExplainer} 
%can predict cancer types 96.25\% of the cases correctly, which 
slightly outperforms the approach by Boyu et al.~\cite{lyu2018deep} but 6.5\% better than the approach by Yuanyuan et al.~\cite{li2017comprehensive}. Further, \emph{OncoNetExplainer} can improve the false prediction rate for the READ, UCS, ESCA, and CHOL tumor samples. In particular, against 35\%, 81\%, 77\%, and 56\% of the correctly predicted cases by~\cite{lyu2018deep}, our approach can predict 88.74\%, 87.26\%, 89.56\%, and 84.55\%~(in cyan) of the same cases correctly.
Although \emph{OncoNetExplainer} performs slightly worse than~\cite{lyu2018deep} at classifying  BRCA, THCA, and PRAD~(in red), it is more consistent for the majority of cancer types and likely to perform more stably on new GE data. 
\emph{OncoNetExplainer} provides both pre-model~(GBT) and post-model interpretation~(CNN and VGG16), whereas~\cite{lyu2018deep} provides only the post-model interpretability. 
some other studies also used GE data~\cite{zhang2016classification,elsadek2018supervised} for the cancer prediction.
%in which an incremental feature selection method was applied to extract GE profiles to train ML models. They managed to achieve an accuracy up to 85\%. % by Zhang et al. and Elsadek et al., respectively). 
However, since GE data from the PCA project had more samples, a one-to-one comparison with these studies was not viable. 
%Further, full convergence to best hyperparameters is often not necessary because a best-fitting model will perform better on unseen data. 

\section{Conclusion and outlook}
\label{conclusion}
In this paper, we proposed \emph{OncoNetExplainer}, an explainable method for the prediction of cancer types based on GE data. Our approach is based on GradCAM++ with CNN and VGG16 networks, and SHAP-based GBT model. Experiment results show that GE is useful for predicting cancer types with high confidence giving an accuracy of up to 96.25\%.
We also attempted to provide a more human-interpretable explanation by showing statistically significant biomarkers. These analyses are further validated with scientific literature~\cite{lawrence2014discovery}, which confirms that the identified genes are biologically relevant.

%However, several other factors have hindered this research, e.g., lack of enough training samples and single modality, i.e., only GE data used for making the predictions and generating the explanations. 
%The latter reason caused the network to produce high training and validation errors during the training phase, giving lower accuracy during the inferencing phase. 
Although we attempted to open the CNN and VGG16 black-box models through biomarker validation and feature ranking,
our approach is mostly post-hoc in that the explainability is based on test cases and results similar to layer-wise relevance propagation. 
Several further factors have hindered this research: i) lack of enough training samples, ii) lack of biological pathways analysis, and iii) since multiple factors are involved in cancer diagnosis~(e.g., estrogen, progesterone, and epidermal growth receptors in BRCA), AI-based diagnoses might not be trustworthy solely based on a single modality, which demands the requirements of multimodal features of DNA methylation, GE, miRNA expression, and CNVs data.

In the future, we intend to extend this work by: i) alleviating more samples by combining genomics data from ICGC and COSMIC to train a multimodal architecture, ii) improving the explanations about the predictions using an ante-hoc approach by seeding explainability into the model from the beginning. In particular, we will focus on multimodality with reversed time attention model and Bayesian deep learning~\cite{choi2016retain}. 
\bibliographystyle{IEEEtran}
\bibliography{reference.bib}

\end{document}